\documentclass[twoside,leqno, twocolumn]{article}

\usepackage[english]{babel}
\usepackage[letterpaper]{geometry}

\usepackage{ltexpprt}

\usepackage{amsmath}
\usepackage{graphicx}
\usepackage[colorlinks=true, allcolors=blue]{hyperref}
\usepackage{amsfonts}
\usepackage{subcaption}
\usepackage{bbm}
\usepackage{tabularx}
\usepackage[switch]{lineno}

\DeclareCaptionFormat{custom}
{\textbf{#1#2}{\small #3} }

\title{A Self-Supervised Algorithm for Denoising Photoplethysmography Signals for Heart Rate Estimation from Wearables}

\author{
    Pranay Jain
    \thanks{Department of Computer Science, Duke University, USA ({pranay.jain455@duke.edu}).}
    \and Cheng Ding \thanks{Department of Biomedical Engineering, Georgia Institute of Technology \& Emory University, USA.}
    \and Cynthia Rudin \thanks{Department of Computer Science, Duke University, USA.}
    \and Xiao Hu\thanks{Professor and Asa Griggs Candler Chair of Nursing Data Science, Associate Director | Center for Data Science, Nell Hodgson Woodruff School of Nursing, Associated Faculty of Biomedical Informatics, School of Medicine, Associated Faculty of Computer Science, College of Arts and Sciences, Emory University, USA.}}

\date{}

\begin{document}
\maketitle
\begin{abstract}
Smart watches and other wearable devices are equipped with photoplethysmography (PPG) sensors for monitoring heart rate and other aspects of cardiovascular health. However, PPG signals collected from such devices are susceptible to corruption from noise and motion artifacts, which cause errors in heart rate estimation. Typical denoising approaches filter or reconstruct the signal in ways that eliminate much of the morphological information, even from the clean parts of the signal that would be useful to preserve. In this work, we develop an algorithm for denoising PPG signals that reconstructs the corrupted parts of the signal, while preserving the clean parts of the PPG signal. Our novel framework relies on self-supervised training, where we leverage a large database of \textit{clean} PPG signals to train a denoising autoencoder. As we show, our reconstructed signals provide better estimates of heart rate from PPG signals than the leading heart rate estimation methods. Further experiments show significant improvement in Heart Rate Variability (HRV) estimation from PPG signals using our algorithm. We conclude that \textit{our algorithm denoises PPG signals in a way that can improve downstream analysis of many different health metrics from wearable devices}.
\end{abstract}



\maketitle

\section{Introduction}
Photoplethysmography (PPG) is a non-invasive optical measurement technique that provides vital information about the cardiovascular system. PPG-enabled devices consist of an optical sensor that measure volumetric variations of blood circulation as PPG signals.  Modern PPG-enabled devices include a variety of technologies such as fingertip-based pulse oximeters, forehead and earlobe-based PPG sensors, and most commonly, wrist-worn smart watches \cite{CastanedaPPGApplications}. PPG monitoring can enable early detection of serious heart conditions that otherwise might go undetected \cite{allen-cardiovascular-ppg, Pereira2020Jan}. A key application of PPG is the estimation of \textit{heart rate} (HR).

PPG is limited by its susceptibility to \textit{noise artifacts}, including motion artifacts (MA) caused by body movements, and artifacts arising from environmental factors like ambient light, sweat, pressure, etc$.$ \cite{sanudo}.  In order to ensure accuracy of HR estimates and robust diagnosis of medical conditions, it is essential to mitigate such artifacts. Methods that address this limitation for prediction of HR from PPG signals can be broadly categorized into two types. The first type estimates HR directly from the signals despite the presence of artifacts \cite{deep-ppg, temko, cornet, ppgnet, PPNet}. The second type of method attempts to extract, denoise or reconstruct a clean signal from the noise-corrupted signal \cite{kalman, deepheart, adaptive-filter-wu, vlsi-wavelet}. These methods output a cleaned signal that can potentially be used for multiple downstream tasks, including HR estimation, which makes them more generally useful. However, methods of the second type tend to perform relatively poorly at the task of HR estimation as compared to the specialized HR estimation models. Furthermore, these approaches reconstruct the entire PPG signal, even if most of the signal may already be artifact-free. This may potentially distort the original signal and cause loss of morphological information even in the useful parts of the signal. Ideally, we would like to have a method that denoises only the noisy part of the signal -- preserving the valuable information in the uncorrupted part -- and provides a clean signal that can be used for accurate HR estimation and for other downstream tasks. That is the focus of the present work.

In this paper, we present a novel method for reconstructing clean PPG signals from noisy signals. Our method outperforms the state-of-the-art for HR estimation and generalizes across datasets better than other approaches. It preserves the useful segments of the PPG signals that are uncorrupted, and only reconstructs the corrupted sections. This is achieved by decoupling the tasks of artifact detection and removal. We apply an  artifact-detection algorithm to remove artifacts from the signal, and then use a denoising autoencoder to reconstruct the signal only in the regions where artifacts were removed. The denoised signal is then used for HR estimation using band-pass filtering and peak detection. This way, our reconstructed signals are more faithful to the truth and more useful for downstream tasks.

An interesting aspect of our approach is the way it leverages publicly available data. There are two sufficiently large and complex public PPG datasets, PPG-DaLiA \cite{deep-ppg} and the Stanford dataset \cite{stanford-data}. PPG-DaLiA records subjects in a wide variety of settings (e.g., walking, driving, eating, cycling). It is a relatively small but high-quality dataset that contains an external means of extracting ground truth HR using electrocardiogram (ECG) signals that are simultaneously recorded and are relatively free from noise. On the other hand, the Stanford dataset is much larger, but has only PPG signals and no external way to assess ground truth HR. Deciding how to best leverage these datasets was a challenge. We chose to use PPG-DaLiA only for out-of-sample testing purposes, since it has ECG for ground truth and thus can provide an honest assessment of HR. We leverage the large collection of real-world clean PPG signals from the Stanford dataset to devise a self-supervised training methodology that is able to reconstruct realistic clean signals.


\begin{figure}[t]
    \centering
    \includegraphics[width=0.5\textwidth]{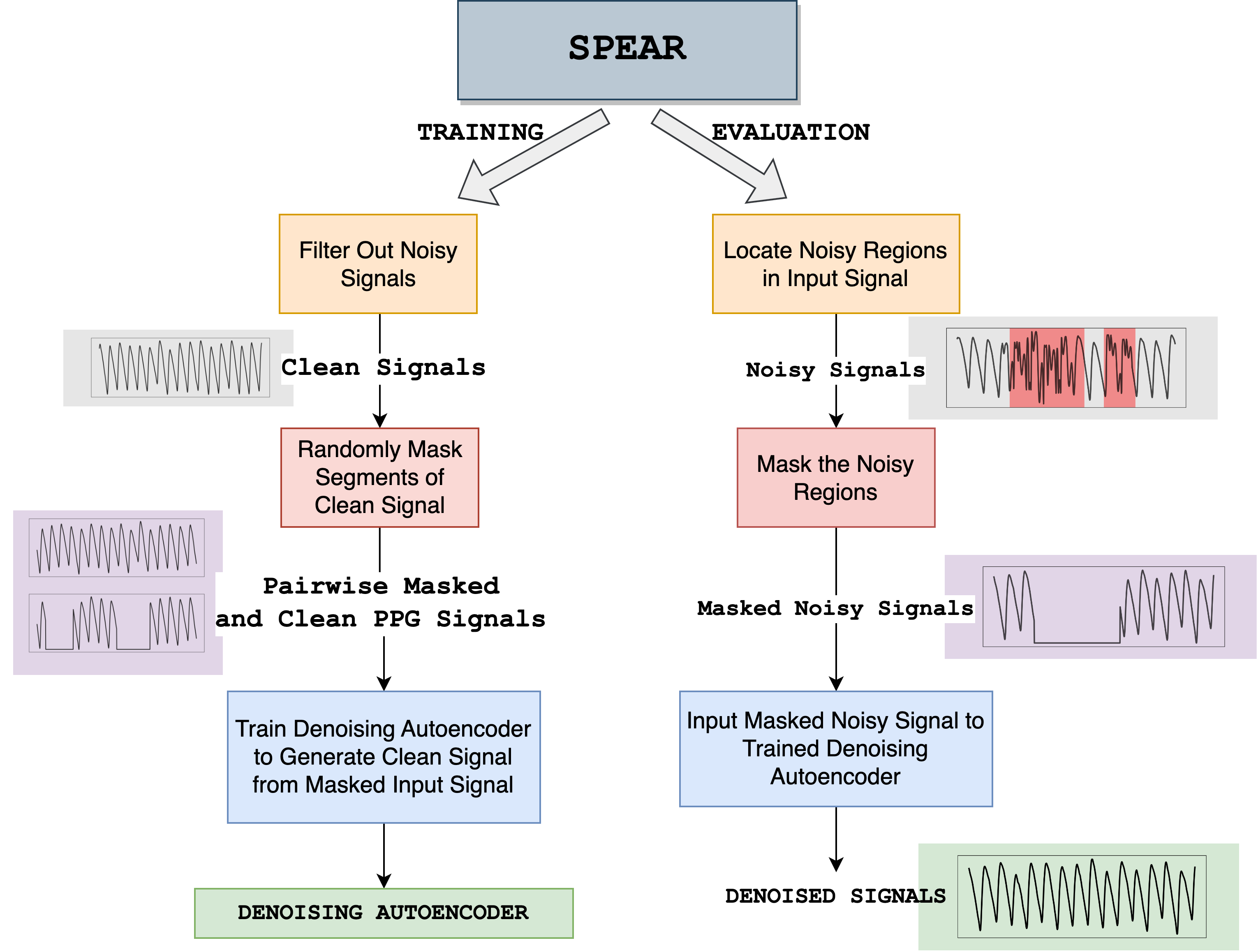}
    \caption{A schematic diagram of the proposed algorithm \texttt{SPEAR}. The left branch outlines the self-supervised training procedure for the denoising autoencoder. The right branch shows how a corrupted signal is denoised by \texttt{SPEAR} using the trained denoising autoencoder and an off-the-shelf artifact detection tool. }
    \label{fig:SpearDiagram}
\end{figure}

Our method is called \textbf{\texttt{SPEAR}} -- Self-supervised PPG Erase Artifacts and Reconstruct -- and it is a novel algorithm for denoising PPG signals. \texttt{SPEAR} learns to denoise PPG signals using the following novel training and evaluation paradigm, outlined in Figure \ref{fig:SpearDiagram} (1) \textit{Removal of segments with artifacts} using an artifact-detection algorithm, leaving only clean signal, (2) \textit{Erasing} random parts of the \textit{clean} signal by setting them to zero, and (3) Training a \textit{denoising autoencoder} to reconstruct these erased parts of the clean signal. The signal is reconstructed in such a way that only the locations that have been erased are reconstructed, and the rest of the signal is unchanged. In this way, given a new noisy signal for testing, our method would (1) apply the artifact-detector, (2) \textit{erase} the artifacts, and (3) \textit{reconstruct} the missing pieces to form a clean signal that can be used for downstream tasks. Since it has learned to reconstruct from \textit{clean} PPG signals in training, it will reconstruct clean signals during testing. We estimate HR using band-pass filtering and peak detection; this type of basic method works precisely because the PPG signal is now clean. 

\texttt{SPEAR} outperforms state-of-the-art methods at the task of heart rate estimation and generalizes well to multiple datasets. Our experiments reveal first that non-machine learning techniques based on signal processing do not tend to perform well for HR estimation in general, and that supervised deep-learning methods tend to perform well on the dataset they were trained on and not beyond that. \texttt{SPEAR} performs about as well on the PPG-DaLiA test set as deep learning methods trained on the same dataset, even though \texttt{SPEAR} is trained on the Stanford dataset. On a hold-out test set from the Stanford dataset, our method outperforms all other methods, substantially outperforming deep learning approaches trained on PPG-DaLiA. Most importantly, \textit{the fact that SPEAR produces clean PPG signals allows the results to be used for downstream tasks} beyond heart rate estimation. We also investigate heart rate variability (HRV) estimation as a downstream task here, showing that HRV estimation is significantly improved using our denoised PPG signals. 
\section{Related Works}


\textbf{Artifact Reduction Techniques.} Signal processing methods have been used to reduce artifacts in PPG signals. Specifically, discrete wavelet transforms \cite{vlsi-wavelet}, adaptive filtering \cite{adaptive-filter-wu, RLS-filter, heartBEAT} and independent component analysis (ICA) \cite{cICA} have been used to perform signal denoising. Salehizadeh et al$.$ \cite{SpaMa} perform sliding window-based signal reconstruction using spectral filtering. A limitation of signal processing approaches is that their performance is dependent on heuristic thresholds and parameters. Reiss et al$.$ \cite{deep-ppg} demonstrate that state-of-the-art signal processing techniques perform poorly on a larger and more comprehensive dataset (PPG DaLiA \cite{deep-ppg}) compared to the (very small) IEEE dataset \cite{troika} that they have been tested on.

Recent works have also introduced deep learning based approaches for this problem. Lee et al$.$ \cite{BRDAE} use a bidirectional recurrent autoencoder for PPG denoising trained on handpicked clean PPG signals.  DeepHeart \cite{deepheart} uses a denoising convolutional network followed by spectrum analysis-based calibration to perform HR estimation. In this approach, signal reconstruction is performed for small overlapping time windows; as a result, the reconstructed clean signals cannot easily be joined together to get a continuous long signal. The survey paper by Mishra et al$.$ \cite{denoise-survey} provides a comprehensive review of the literature on PPG denoising techniques.

A challenge with denoising is the availability of noisy-clean signal pairs in the data, which is required for supervised learning. This is hard to obtain with PPG signals because it is not possible to record clean and noisy signals synchronously while performing certain activities. Workarounds are typically used to overcome this challenge, such as generating fake noisy signals by adding simulated noise to clean signals \cite{BRDAE}. We will discuss how we overcome this challenge using self-supervised learning in Section \ref{method}. In section \ref{ResultsSection}, we demonstrate our method's superior performance to the training approach of simulating PPG signals \cite{BRDAE}.





\textbf{Direct Heart Rate Estimation Without Denoising.} A category of approaches focus on the task of estimating heart rate (HR) directly from noisy PPG signals, without attempting to reconstruct or denoise the signal. Signal processing techniques including Wiener Filtering \cite{temko}, Least-Means Square Adaptive Filtering \cite{shaeck-hr} and TROIKA \cite{troika} utilize accelerometer data and analyze the signals in the frequency domain. Deep learning has also been utilized for HR estimation, most commonly as a supervised learning task where ground-truth HR labels are obtained from synchronous Electrocardiogram (ECG) signals. DeepPPG \cite{deep-ppg} and PPGnet \cite{ppgnet} use deep convolutional networks to predict heart rate from noisy PPG signals. CorNET \cite{cornet} uses a combination of CNNs and LSTMs to predict HR from single-channel PPG in patient-specific models. PP-Net \cite{PPNet} also uses CNNs and LSTMs for HR estimation using single-channel PPG. These approaches tend to outperform the denoising approaches for HR estimation, but do not output a denoised signal that can be utilized in downstream analysis. 


\textbf{Heart Rate Variability (HRV) Estimation from PPG. } HRV measures the fluctuation in the time intervals between adjacent heartbeats \cite{hrv-overview-shaffer}. HRV is used to investigate the sympathetic and parasympathetic function of the autonomic nervous system \cite{hrv-overview-shaffer} and has many important applications including predicting risk of stroke \cite{stroke-hrv}, detecting arrhythmia \cite{arrhythmia-hrv} and guiding training for athletes \cite{fitness-hrv}. Studies that use PPG signals to estimate HRV focus only on clean signals obtained from subjects at rest \cite{ppg-hrv}. A recent study showed poor performance of HRV estimation from PPG signals under free-living conditions \cite{hrv-real-world}. Denoising PPG signals can improve the accuracy of HRV monitoring in real-world conditions, as we will show.

\textbf{Artifact Detection Techniques.} A separate body of work aims to locate the presence of noise artifacts in PPG signals \cite{chong-artifact-detection, Goh2020Nov, segade}. One approach  \cite{Goh2020Nov} divides the signal into sliding windows and uses CNNs to classify whether each window contains an artifact. Another work \cite{segade} approaches this problem as a 1D segmentation problem and uses a convolutional network to classify each location within a 30-second signal segment as to whether it represents an artifact. This allows for detection of noise artifacts on a higher resolution. These approaches only detect the locations of noise artifacts and do not provide further steps on mitigating the artifacts for HR analysis. We utilize the Segade model \cite{segade} as a preprocessing step in \texttt{SPEAR}.

\textbf{Denoising Autoencoders.} Autoencoders are networks that learn to reconstruct their inputs from a latent representation. An autoencoder takes as input a vector $\mathbf{x}$ and maps it to a hidden latent representation $\mathbf{r}$. The resulting latent representation $\mathbf{r}$ is then mapped back to a ``reconstructed'' vector $\mathbf{z}$. The model is trained to minimize the reconstruction error between $\mathbf{x}$ and $\mathbf{z}$. A \textit{denoising autoencoder} is trained to reconstruct a clean input from a corrupted or partially destroyed one \cite{denoise-ae}. For training, the input vector is first partially corrupted, which yields a mapping $ (\tilde{\mathbf{x}}, \mathbf{x})$, where $\tilde{\mathbf{x}}$ is the vector resulting after corrupting $\mathbf{x}$. After obtaining the $ (\tilde{\mathbf{x}}, \mathbf{x})$ mappings, an autoencoder network is trained to reconstruct a vector $\mathbf{z}$ given $\tilde{x}$ as input, such that it minimizes the reconstruction error between $\mathbf{z}$ and $\mathbf{x}$. It is not immediately clear how one would apply a denoising autoencoder to denoise PPG signals from unlabeled data; the novelty of our approach is a framework that incorporates it. 

\section{Method} \label{method}
We propose a self-supervised training approach that requires only a sufficiently large collection of \textit{clean} PPG signals, and does not require synchronous ECG measurements like other approaches \cite{deepheart, ppgnet, PPNet, cornet, deep-ppg}. Using a self-supervised training approach solves the challenge of unavailable noisy-clean training pairs required for supervised learning. This is because the model is trained to reconstruct clean signals from signals where the noise has been \textit{erased} -- it requires no information about the noise artifacts except for where they are located.



\subsection{Training Process for \texttt{SPEAR}.} \label{training}
\begin{figure*}[t]
    \centering
    \includegraphics[width=\textwidth]{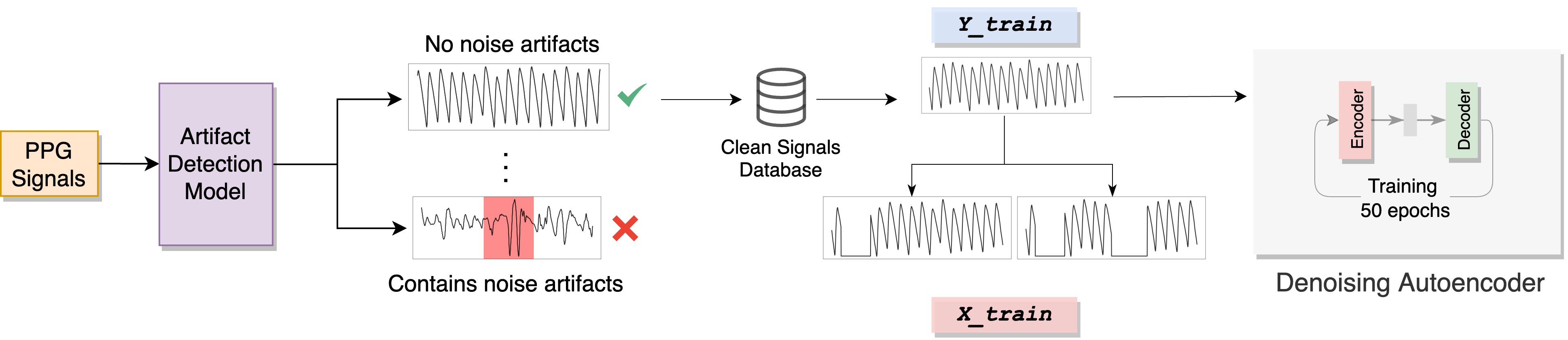}
    \caption{Training: Training Procedure for the Denoising Autoencoder}
    \label{fig:DAEtraining}
    \vspace{2em}
    \includegraphics[width=\textwidth]{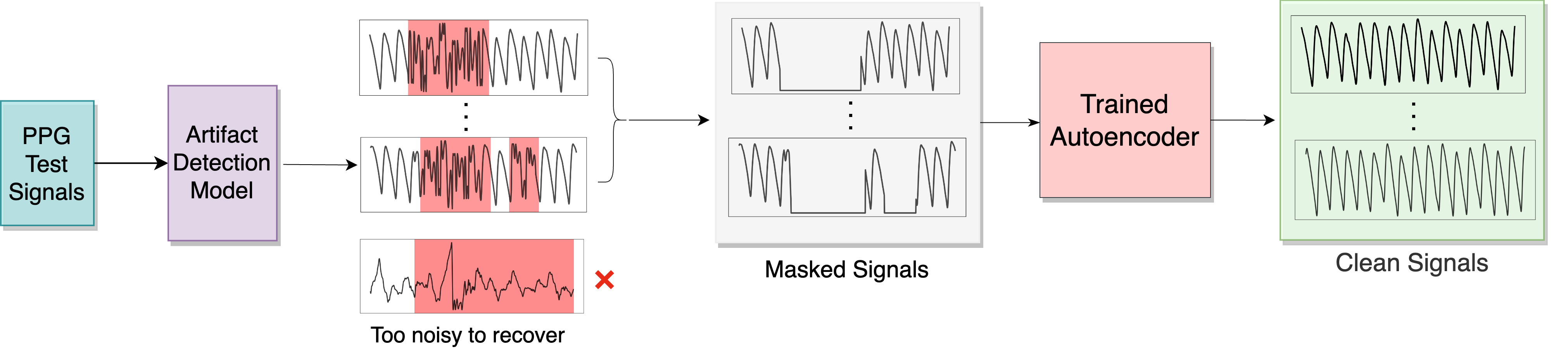}
    \caption{Evaluation: Denoising Test PPG Signals Using the SPEAR Algorithm.}
    \label{fig:DenoisingAlgo}
\end{figure*}

In this subsection, we outline the methodology of training \texttt{SPEAR}'s specialized denoising autoencoder. Figure \ref{fig:DAEtraining} summarizes the training procedure. The process for denoising a new signal will be discussed afterwards and in Figure \ref{fig:DenoisingAlgo}.

\subsubsection*{Data.}
The Stanford training set was used for \textit{training} the reconstruction model because it contains a large number of clean PPG signals. The Stanford test set and PPG DaLiA test sets were used for \textit{evaluating} the model. The PPG DaLiA test set was used, instead of testing on the complete PPG DaLiA dataset, to ensure fair comparison with the baseline methods that were trained on the PPG DaLiA training set.

\subsubsection*{Selecting Clean PPG Signals.} \label{preprocessing}
The first step to prepare the training data is selecting clean PPG signals. The Segade model \cite{segade} was used for this purpose. 
Given a 30-second PPG signal as input, Segade  predicts the regions within the signal that are corrupted by noise artifacts. Any signal that was determined to have no corrupted regions was deemed as a clean signal.


\subsubsection*{Preparing Training Data.}
In the next step, a denoising autoencoder (DAEs) \cite{denoise-ae} is given a partially corrupted signal as input and trained to recover the original signal. Training a DAE requires element-wise pairs of signals $(X, \tilde{X})$ where $\tilde{X}$ is a partially corrupted or destroyed version of $X$. To prepare the training dataset, each clean PPG signal was partially erased at one or more continuous sub-segments. More precisely, we create mask vectors $\{M_i\}$ of the same dimension as $X$: $M_i$ has the value 1 everywhere except one or more continuous patches, where it is 0. We start with $M_i = \mathbf{1}$ (vector of all 1's). A starting point $s_i$ and a patch length $l_i$ were randomly selected and the values $M[s_i : s_i + l_i]$ were set to 0. The patch lengths were varied between 1 and 15 seconds, and the number of patches in a mask was set to either 1 or 2. For each signal, 10 such masks were generated, each yielding a new training sample $(X \odot M_i, X)$, where $X$ is the clean signal, and $\odot$ is the element-wise (Hadamard) product.  

\subsubsection*{Model Architecture and Training Parameters.}
The denoising autoencoder consists of an encoder network that maps the input signal to its latent space, and a decoder network that reconstructs the clean signal from the latent space input. The encoder network consists of 4 convolutional layers, each followed by a ReLU activation and batch normalization.  The decoder network consists of 4 Transpose Convolution-ReLU-BatchNormalization blocks. The fourth block is followed by a convolutional layer with a sigmoid activation that outputs the reconstructed signal. The encoder receives input in the dimension $(N, 1920)$, where $N$ is the number of 30-second signals sampled at 64Hz. Details on the exact model architecture are provided in the supplementary material.

Loss was computed as the Root Mean Square Error (RMSE) between the original clean and the reconstructed signals. The model was optimized using the Adam optimizer \cite{adam}, trained over 50 epochs.





\subsection{Using \texttt{SPEAR} to Denoise PPG Signals.}
In Section \ref{training}, we discussed the denoising autoencoder used in our algorithm. In this section, we provide an end-to-end framework for denoising PPG signals and estimating HR using \texttt{SPEAR}. This process is illustrated in Figure \ref{fig:DenoisingAlgo}.

\subsubsection*{Step 1: Artifact Detection.}
The first step in the signal denoising algorithm is to locate the noise artifacts. The Segade model is again used for this purpose. Similar to the preprocessing defined in Section \ref{preprocessing}, first the signals are fitted to 30 second segments sampled at 64 Hz. The signals are normalized to the $[0,1]$ range using min-max normalization. The Segade model receives these signal segments as input and outputs predictions for the noise artifact locations. A threshold of 0.5 is applied to determine the binary classification labels. 



Signals that are excessively corrupted beyond recovery are discarded. We consider a $75\%$ threshold of cumulative duration, beyond which a 30-second signal is considered unrecoverable. Intuitively, it is reasonable to assume that it may be futile to recover a 30-second signal where 27 seconds are corrupted by noise artifacts, since too much of the information is lost to noise.

\subsubsection*{Step 2: Signal Reconstruction.}
The remaining signal is used as input for the denoising autoencoder model. The locations where Segade output has a classification label of 1 (indicating the presence of a noise artifact) are \textit{erased} in the signal (set to 0). Let $\mathbf{X}_{\textrm{seg}}$ be the signals obtained after discarding the excessively corrupted signals in the previous step. Similarly define $\mathbf{Y}_{\textrm{seg}}$ as the binary classification of noise artifacts by Segade corresponding to $\mathbf{X}_{\textrm{seg}}$. Then the input to the denoising autoencoder can be obtained as:
$$\mathbf{X}_{\textrm{in}} = \mathbf{X}_{\textrm{seg}} \odot (\mathbbm{1} - \mathbf{Y}_{\textrm{seg}}).$$

Here, $\mathbbm{1}$ is a matrix comprising of 1 at all positions. The denoising autoencoder receives as input the matrix $X_{\textrm{in}}$, and outputs $Y_{\textrm{out}}$ which consists of clean PPG signals. Intuitively, since the denoising autoencoder was trained to recover clean signals from partially corrupted ones, $Y_{\textrm{out}}$ consists of clean PPG signals.

\subsubsection*{Step 3: Post-Processing and Heart Rate Detection.}
The final step is to post-process the clean signals in $Y_{\textrm{out}}$. First, the denoised signal is \textit{merged} with the original signal $X_{\textrm{in}}$, in that the artifact-corrupted regions of $X_{\textrm{in}}$ are replaced with the reconstructed regions in $Y_{\textrm{out}}$. This can be obtained as:
$$X_{\textrm{denoised}} =  X_{\textrm{in}} \odot (\mathbbm{1} - Y_{\textrm{seg}}) +  Y_{\textrm{out}} \odot Y_{\textrm{seg}},$$

where $Y_{\textrm{seg}}$ is the binary classification by Segade and contains 1's in the noise-corrupted regions and 0 otherwise. After this, $X_{\textrm{denoised}}$ can be utilized for downstream analysis tasks.


 To perform HR estimation, a bandpass filter with low-end cutoff of 0.9 Hz and a high-end cutoff of 5 Hz is applied, and the signal is re-normalized to the $[0,1]$ range using min-max normalization. Since the resulting signal is clean, a simple peak-detection algorithm can be applied to perform HR estimation. We use the  peak-detection algorithm by \cite{elgendi-peak-detection} which is implemented in the HeartPy python package \cite{heartpy}.

\section{Experimental Setup} \label{ExperimentalSection}
In this section, we describe the datasets, baselines, experimental setup and evaluation metrics for heart rate estimation.

\subsection{Datasets}
Two datasets were used in this study: the Stanford Dataset \cite{stanford-data} and PPG DaLiA Dataset \cite{deep-ppg}; both datasets consist of PPG recordings collected from a wrist-worn device sampled at 64Hz.  Table \ref{data-properties} compares the main properties of these two datasets. The Stanford dataset was selected because it contains the largest publicly available collections of PPG signals from wrist-worn wearables. PPG DaLiA was used to train baselines because it contains synchronously recorded ECG and accelerometer data, which are required by most existing methods for HR estimation.

The Stanford dataset is divided into training, validation and test sets with no subject overlap. The training set was used for training the denoising autoencoder. The validation set was used for hyperparameter tuning. The test set was used for testing the performance of \texttt{SPEAR} in comparison with baselines. The PPG DaLiA dataset \cite{deep-ppg} was used for out-of-sample testing and comparison against baselines. PPG DaLiA was selected because it has a comprehensive data collection protocol from subjects of different ages while performing a variety of daily activities such as walking, cycling, driving and working on a laptop, among others.

\begin{table}[h]
\setlength{\tabcolsep}{2pt}
\small
\begin{center}
\begin{tabularx}{\columnwidth}{
  | >{\centering\arraybackslash}X 
  | >{\centering\arraybackslash}X 
  | >{\centering\arraybackslash}X | }

\hline  
 \textbf{Property} & PPG DaLiA & Stanford \\
 \hline
 \textbf{PPG} & Available & Available  \\
 \textbf{ECG} & Available & Unavailable  \\
 \textbf{Accelerometer} & Available & Unavailable \\
\textbf{No. of subjects} & 15 & 149 \\
 \textbf{No. of train samples (clean)} & $3.4$k (233) & 62k (7.4k) \\
 \textbf{No. of test samples} & 872 & 6.7k \\
 \hline
\end{tabularx}
\caption{Properties of the two datasets used in our study. A sample corresponds to a PPG signal of length 30 seconds.}
\label{data-properties}
\end{center}
\end{table}

\subsection{Baselines}
In this section, we introduce the 6 state-of-the-art baseline methods that were used for comparison. Baselines 1 and 2 use signal processing techniques and were chosen as they achieved the best performance on the IEEE Signal Processing Cup data \cite{troika}. Baselines 3-6 use deep learning for HR estimation without denoising; the models are based on the works of \cite{PPNet, deep-ppg, cornet, ppgnet}. Although none of these studies have publicly available code, these approaches were chosen for their performance and ease of implementation. Implementations for other baselines \cite{deepheart, troika, BRDAE, cICA, adaptive-filter-wu, RLS-filter, SpaMa} are not publicly available and not reproducible.


\begin{itemize}
    \item \textbf{Baseline 1: Wiener Filtering and Phase Vocoder (WFPV).} This baseline is based on \cite{temko}. It uses the three-axis accelerometer signals to estimate the noise signature and applies a Wiener filter to attenuate noise components in the frequency domain. A Phase Vocoder is used to estimate HR. 

    \item \textbf{Baseline 2: Kalman Filtering.} This baseline is based on \cite{kalman} and is a signal processing technique that produces a reconstructed PPG signal over small time windows. It performs signal decomposition over PPG and three-axis accelerometer signals and performs clean PPG reconstruction based instead on the degree of correlation of PPG with accelerometer signals. Kalman smoothing is used for HR estimation from the reconstructed signal. 
    \item \textbf{Baseline 3: \texttt{CNN\_HR\_DaLiA}}. This baseline model uses supervised learning to estimate HR directly from sliding time windows over a noisy PPG signal. Our version of this baseline was based on DeepPPG \cite{deep-ppg} and PPGNet \cite{ppgnet}. The single-channel PPG signals in PPG DaLiA dataset were used for training and HR ground-truth labels were obtained from the synchronously recorded ECG.

    \item \textbf{Baseline 4: \texttt{CNN+LSTM\_HR\_DaLiA}.} This baseline follows the same approach of direct HR estimation on sliding windows as Baseline 3. Our version of this baseline is based on the PP-Net \cite{PPNet} and CorNET \cite{cornet} models. It is trained using the same procedure as Baseline 3; the only difference is in the model architecture. This model uses a combination of convolutional and Long-Short-Term-Memory (LSTM) layers. 
    \item \textbf{Baseline 5: \texttt{CNN\_HR\_Stanford}}. This model is architecturally identical to Baseline 3, but was trained on the Stanford training set. Since the Stanford dataset includes a significantly larger collection of signals, it was important to establish comparisons with baselines trained on similar data as \texttt{SPEAR}. However, since the Stanford dataset does not contain ECG signals for ground-truth labels, the technique of clean signal selection and simulated corruption to generate noisy-clean training pairs was used. Clean signals were selected using the technique defined in Section \ref{preprocessing} and simulated noise artifacts were added using the RRest toolbox \cite{rrest-charlton}: a combination of Frequency Modulation (FM) and Baseline Wander (BW) were added to clean signals, while ensuring that no more than 75\% of the signal is corrupted, to match the maximum degree of corruption expected by \texttt{SPEAR}. HR labels were generated on the clean PPG pairs using \cite{elgendi-peak-detection}'s technique.

    \item \textbf{Baseline 6: \texttt{CNN+LSTM\_HR\_Stanford}.} This model is trained on the Stanford training set using the procedure as Baseline 5. The model uses a combination of convolutional and LSTM layers and has an identical architecture to Baseline 4.

    \item \textbf{Baseline 7: \texttt{DAE\_SimNoise}.} This is a denoising model based on the training approach of \cite{BRDAE}. To train this model, noisy-clean pairs of PPG signals were generated by selecting clean PPG signal segments of duration 30 seconds and adding simulated noise. The noise simulation procedure was similar to Baselines 5-6, where a combination of FM and BW noise were added, while ensuring total corruption is under 75\%. A training dataset was generated from the Stanford data such that the number of samples were roughly equal to \texttt{SPEAR}'s training set. A denoising autoencoder with an identical architecture to \texttt{SPEAR} was trained to denoise the signals.
\end{itemize}

\begin{table*}[h]
\setlength{\tabcolsep}{3pt}
\small
\begin{center}
\begin{tabularx}{\textwidth}{
   >{\centering\arraybackslash}X 
   >{\centering\arraybackslash}X 
   >{\centering\arraybackslash}X 
   >{\centering\arraybackslash}X 
   >{\centering\arraybackslash}X  }

\hline  
 & Method & Training Dataset & PPG DaLiA MAE & Stanford MAE \\
 \hline
 Baseline 1 & \texttt{WFPV} & - & 21.07 & -  \\
 Baseline 2 & \texttt{Kalman Filtering} & - & 20.20 & -  \\
 Baseline 3 & \texttt{CNN\_HR\_DaLiA} & PPG DaLiA & 6.60 & 11.39 \\
 Baseline 4 & \texttt{CNN+LSTM\_HR\_DaLiA} & PPG DaLiA & 4.17 & 7.24 \\
 Baseline 5 & \texttt{CNN\_HR\_Stanford} & Stanford & 12.97 & 5.26 \\
 Baseline 6 & \texttt{CNN+LSTM\_HR\_Stanford} & Stanford & 9.03 & 4.06 \\
  Baseline 7 & \texttt{DAE\_SimNoise} & Stanford & 8.44 & 5.93 \\
 \textbf{Our Algorithm} & \texttt{SPEAR} & Stanford & \textbf{5.36} & \textbf{3.18} \\
 \hline
\end{tabularx}
\caption{Experiment Results: Mean Absolute Error (MAE) on heart rate estimation. Columns 1-2  lists the methods used for HR estimation, Column 3 lists the training dataset for the corresponding method. Column 4 lists MAE results from the PPG DaLiA experiment (\ref{experiments} - \ref{dalia-expt}) and Column 5 lists the MAE results from the Stanford experiment (\ref{experiments} - \ref{stanford-expt}). }
\label{MAEResults}
\end{center}
\end{table*}



\subsection{HR Estimation Experiments Setup} \label{experiments}

\begin{enumerate}
    \item \textbf{PPG DaLiA Experiment:} \label{dalia-expt} The PPG DaLiA dataset was divided into a training and test set. The test set contains signals from subjects 1, 14 and 15. The signals were first split into non-overlapping 30-second segments. Segments that were more than 75\% corrupted are discarded (same as our method). The corresponding ECG as well as the three accelerometer signals were similarly segmented. The accepted signals were joined into one longer signal and used for testing the baselines. For Baselines 1 and 2, the continuous subject-wise signals were used as input. For Baselines 3-6, the signals were segmented into 8-second overlapping windows (with an overlap of 6 seconds). Heart Rate estimation was performed on the PPG signals using the ECG for ground-truth labels. 

    \item \textbf{Stanford Dataset Experiment:} \label{stanford-expt} The Stanford test set was used for this experiment.  Since this dataset does not contain ECG for ground truth measurement, we utilized the clean signals as the source of ground truth. We introduce simulated noise in the clean signals similar to the training procedure defined for Baseline 5 using the RRest package \cite{rrest-charlton}. This produced clean-noisy signal pairs. Ground-truth HR was computed on the clean signal using a peak detection algorithm \cite{elgendi-peak-detection}. Baselines 1 and 2 could not be evaluated on this dataset since they require 3-axis accelerometer data as input, which is not available in the Stanford data. 
\end{enumerate}

\subsection{Heart Rate Variability (HRV) Estimation}
In this experiment, we estimated Heart Rate Variability (HRV) from PPG signals. The goal of this experiment was to evaluate whether denoised signals produced by \texttt{SPEAR} can be utilized for downstream tasks and provide improvements over existing methods. Several metrics are used to measure HRV. In our experiments, we focus on two metrics: SDNN is the standard deviation of the inter-beat intervals measured in milliseconds (ms), and RMSSD is the root mean square of the successive differences between normal heartbeats measured in ms \cite{hrv-overview-shaffer}. A review of HRV-capable wearable devices shows that RMSSD and SDNN are the two metrics that are most commonly available on such devices \cite{HRVDevices}. SDNN is generally studied in clinical settings, considered to be the "gold standard" metric for assessing cardiac risk, and used for predicting morbidity and mortality \cite{hrv-overview-shaffer}.  

HRV requires continuous measurement over long time duration; typically, a time window of 5 minutes is used for short-duration estimation \cite{hrv-overview-shaffer}. The full PPG DaLiA dataset was used for evaluation. The signals were segmented into sliding windows of duration 5 minutes and an overlap of 95\%. The two HRV metrics (SDNN and RMSSD) were computed from PPG signals and synchronously recorded ECG (for ground-truth). Mean absolute error was computed between HRV estimates from PPG and HRV ground truth from the corresponding ECG. 

There is a dearth of literature on directly estimating HRV from PPG, and existing denoising approaches  \cite{kalman, deepheart, adaptive-filter-wu, vlsi-wavelet} do not directly work for our HRV estimation task. This is because they perform denoising on short, overlapping (8-second) signal segments for heart rate estimation; they don't offer methods on reconstructing longer signals that can be used for HRV. Consequently, the baselines used for HR estimation could not be adapted for the HRV task, except baseline 7. \texttt{SPEAR}, as well as baseline 7, reconstruct non-overlapping 30-second signals that can be combined into continuous long-term recordings using interpolation. Thus, existing approaches based on simple peak detection can be applied for HRV estimation. For comparison, we perform HRV estimation on four variants of the PPG signals: the original signal from the dataset, the signal after applying bandpass filtering, the denoised signal from \texttt{DAE\_SimNoise} (baseline 7) and the denoised signal from \texttt{SPEAR}.

To estimate HRV from PPG, we adapt the widely used methods of \cite{SinusCor, Morelli2018Apr}. First, a peak detection algorithm \cite{elgendi-peak-detection} detects the R-peaks. Then, a moving filter is applied to remove physiologically implausible peaks. The filtering criterion differs over studies \cite{SinusCor, Morelli2018Apr}; we used a filter based on Inter-Quartile Range, which rejects R-R intervals that lie outside of the IQR of measured interval durations. The HRV metrics (SDNN and RMSSD) were computed using their respective formulae based on R-R intervals. The implementation for these approaches are available in the HeartPy library \cite{heartpy}.


\subsection{Evaluation Metrics.}
The Mean Absolute Error (MAE) is a widely used metric in HR estimation challenges. HR is estimated on signals segmented into sliding time windows of length 8 seconds and an overlap of 6 seconds. When the signals are split into $N$ overlapping windows, we let $\{ \text{BPM}_\textrm{est}(i) \}_{i=1}^N$ be the HR estimated from the PPG windows. For reconstruction techniques, this is estimated from the denoised signal. Similarly, let $\{ \text{BPM}_\textrm{true}(i) \}_{i=1}^n$ be the HR estimated from the ground-truth windows. Then the mean absolute error is: 
$$ MAE = \frac{1}{N} \sum_{i=1}^n \Big | \text{BPM}_\textrm{est}(i) - \text{BPM}_\textrm{true}(i) \Big |. $$

For Heart Rate Variability, the same approach is used, but the HRV metrics are computed over a 5 minute interval, based on the recommended time interval used for HRV \cite{hrv-overview-shaffer}. The MAE is computed on the two HRV metrics, SDNN and RMSSD, using the ECG segments as ground-truth. 

\section{Experimental Results}\label{ResultsSection}

\subsection{Heart Rate Estimation Results.}
Our main result is that the proposed algorithm outperforms the baselines at the task of HR estimation from PPG signals. The results are summarized in Table \ref{MAEResults}. In the Stanford experiment, \texttt{SPEAR} \textbf{\textit{outperforms all baselines and achieves the lowest MAE}}. In the PPG DaLiA experiment, our algorithm has a lower MAE than every baseline except the \texttt{CNN+LSTM\_HR\_DaLiA} model (which was itself trained on PPG-DaLiA), though the difference in MAE is not big. This shows that \texttt{SPEAR}'s \textbf{\textit{out-of-distribution performance is on par with fully supervised approaches trained on the same distribution.}}

The deep learning-based HR estimation models (Baselines 3-6) perform better than the signal processing approaches. Furthermore, the models using an architecture utilizing convolutional along with recurrent networks (LSTMs) outperform the purely convolutional networks. That being said, these approaches do a worse job of generalizing to different datasets. The \texttt{CNN+LSTM\_HR\_DaLiA} baseline (trained on the PPG DaLiA training data) performs very well on the PPG DaLiA test set, but does a worse job on the Stanford test data. The \texttt{CNN+LSTM\_HR\_Stanford} baseline (trained on the Stanford training data) has good performance on the Stanford test data, but significantly poorer performance on the PPG DaLiA dataset. In contrast, \textit{\textbf{our approach generalizes well to multiple datasets and has superior performance on both datasets}}. It is also evident that approaches trained on PPG containing simulated noise (baselines 5-7) achieve poorer performance on out-of-distribution data in comparison to \texttt{SPEAR}. This shows that our self-supervised technique of erasing noise artifacts, instead of simulating them, generalizes better to signals that contain noise artifacts under real-world conditions.

\begin{figure}[h]
    \centering
    \includegraphics[width=\columnwidth]{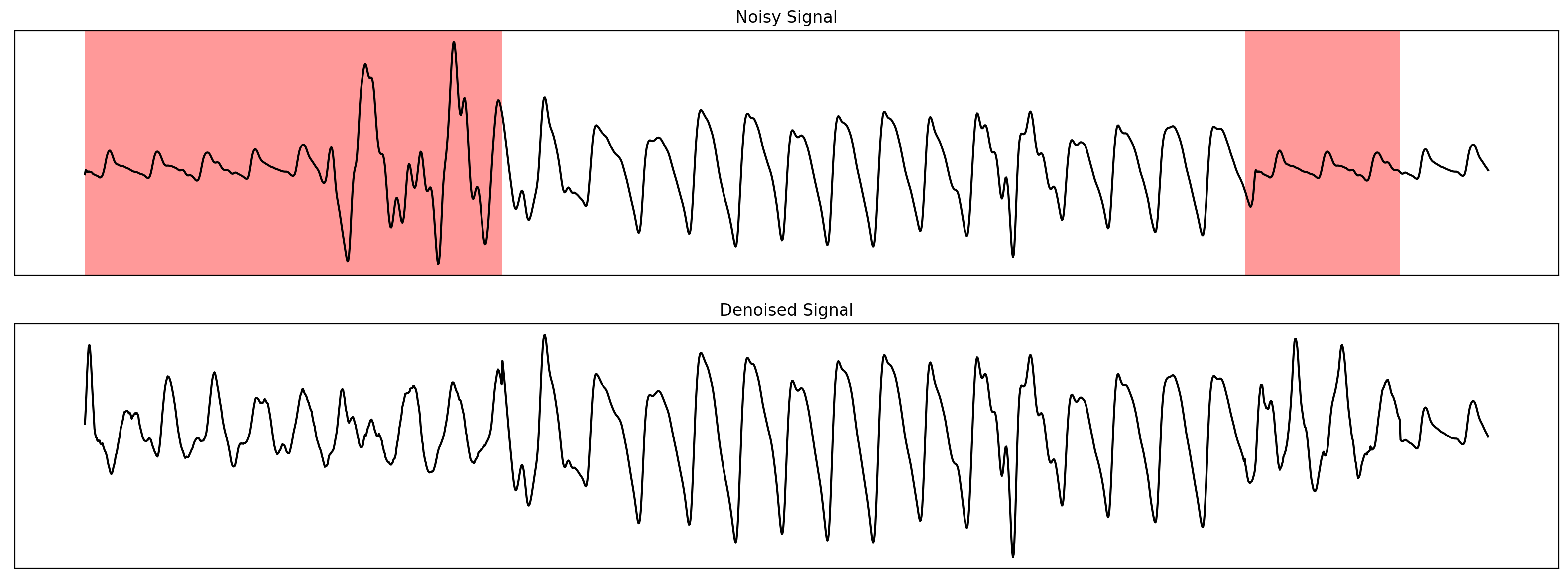}
    \caption{An example of denoising on a 30s signal segment from the Stanford data (sampling rate: 64Hz). Top: Signal corrupted by noise artifacts (highlighted in red). Bottom: The denoised signal produced by \texttt{SPEAR}.}
    \label{fig:stan-denoise}
\end{figure}

\begin{figure}[h]
    \centering
    \includegraphics[width=\columnwidth]{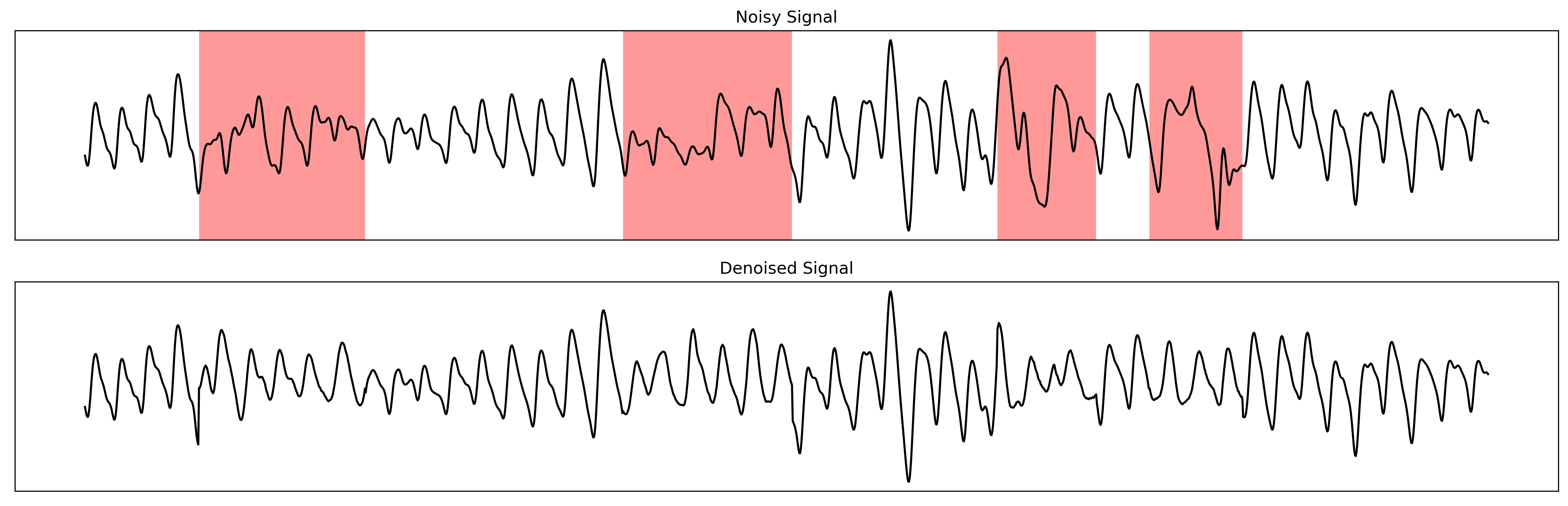}
    \caption{An example of denoising on a 30s signal segment from the PPG DaLiA data (sampling rate: 64Hz).}
    \label{fig:dalia-denoise}
\end{figure}

\begin{figure}
    \centering
    \begin{subfigure}[b]{\columnwidth}
        \includegraphics[width=\columnwidth]{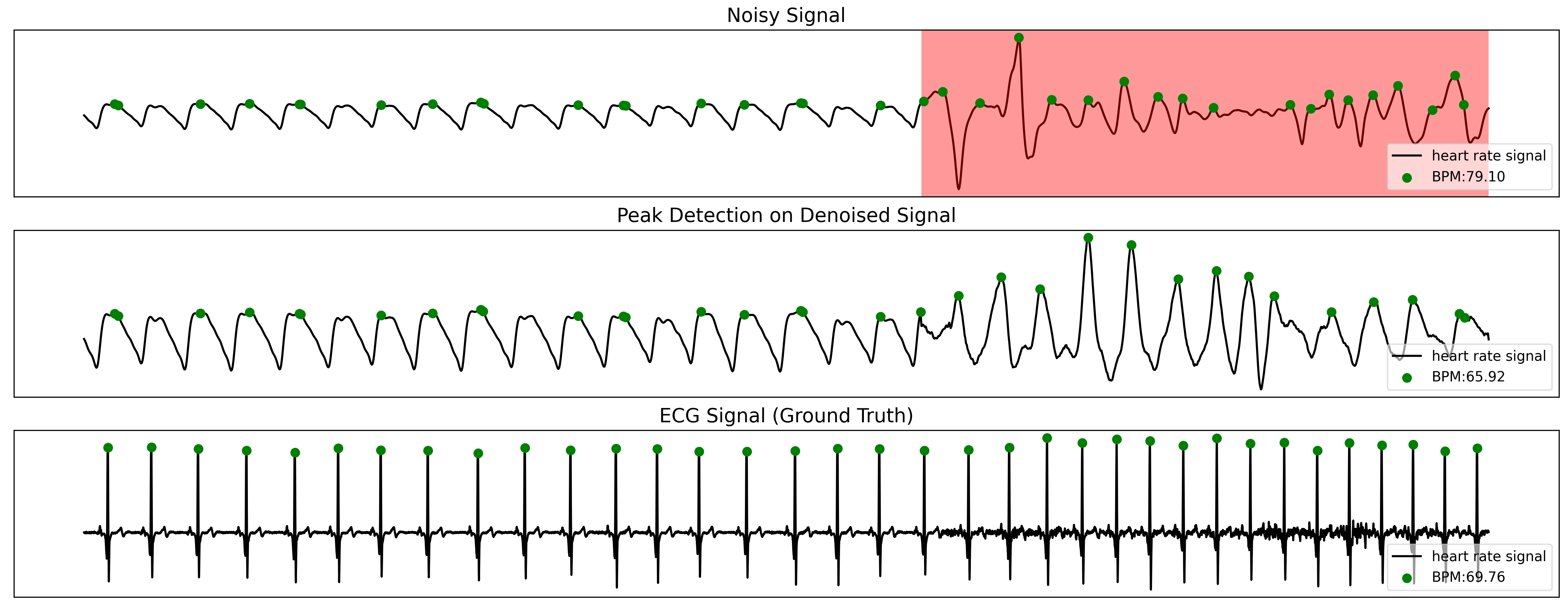}
        \caption{$HR_{\text{true}} = 69.8$ bpm. $HR_{\text{noisy}} = 79$ bpm. $HR_{\text{denoised}} = 65.9$  bpm.}
    \end{subfigure}

    \begin{subfigure}[b]{\columnwidth}
        \includegraphics[width=\columnwidth]{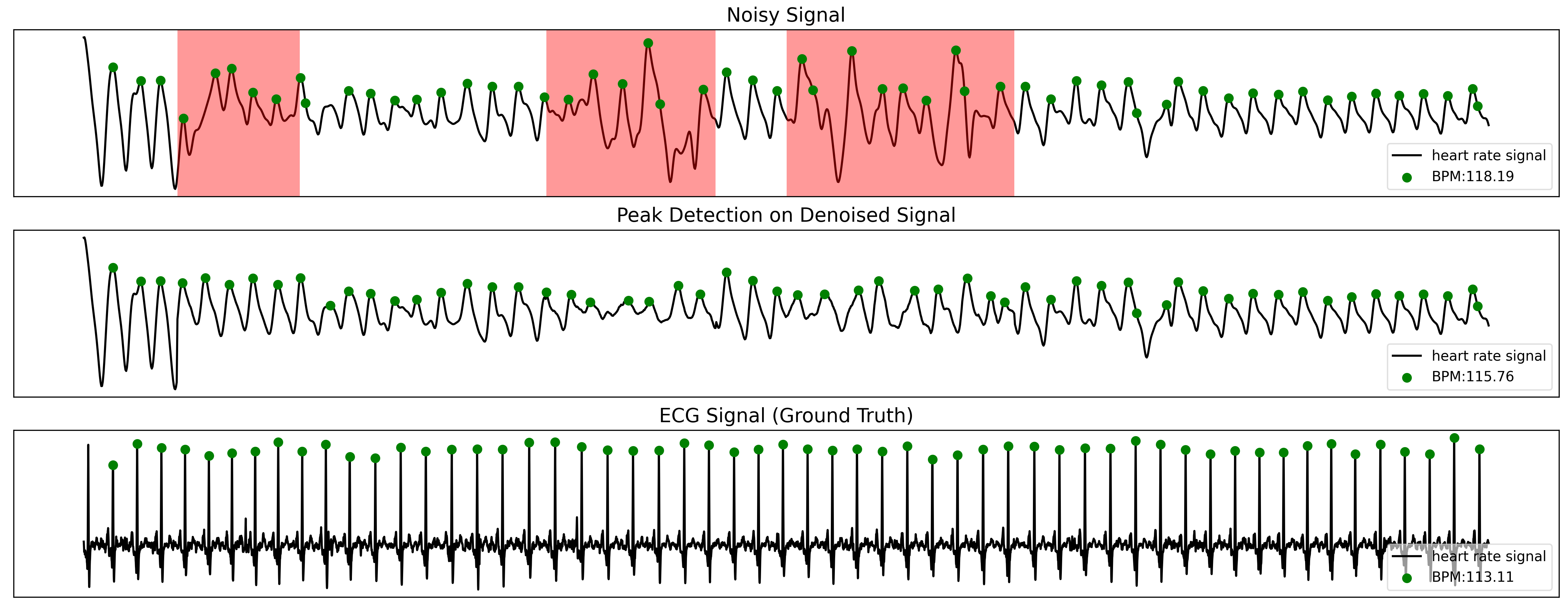}
        \caption{$HR_{\text{true}} = 113$ bpm. $HR_{\text{noisy}} = 118$ bpm. $HR_{\text{denoised}} = 115$  bpm.}
        
    \end{subfigure}
    \caption{Two examples of 30s signal segments from the PPG DaLiA dataset. In both examples, the first signal is the noisy PPG signal (heart rate $HR_{\text{noisy}}$), with the red regions highlighting the detecting noise artifacts. The second signal is the denoised signal produced by \texttt{SPEAR}, with heart rate $HR_{\text{denoised}}$. The third signal is the ground-truth ECG signal with heart rate $HR_{\text{true}}$. The green dots indicate detected R-peaks.}
    \label{fig:SignalsWithPeaks}
\end{figure}

Figures \ref{fig:stan-denoise} and \ref{fig:dalia-denoise} show examples of denoised signals produced by \texttt{SPEAR}. Figure \ref{fig:SignalsWithPeaks} shows two examples from the PPG DaLiA dataset, along with a visualization of peak detection on the noisy and denoised signals. The figure demonstrates denoising under different conditions --- in the first signal, the subject has a normal heart rate, but introduced some motion artifacts, while in the second, the subject has elevated heart rate. 


\subsection{HRV Estimation Results.}
The main result from this experiment is that \textit{\textbf{\texttt{SPEAR}'s denoised signals achieve significant improvements in Heart Rate Variability estimation tasks over existing techniques}}. Table \ref{HRV-Results} shows the mean absolute error in HRV estimation from PPG signals in our experiment. HRV is measured from the inter-beat time intervals in milliseconds; as a result, HRV is highly sensitive to noise because artifacts can cause missed or extra beats, which leads to large errors in the interval measurements. Our results confirm that existing techniques of estimating HRV from PPG signals \cite{SinusCor, Morelli2018Apr} produce large errors compared to the ground-truth. Denoising the signals achieves improved estimates over the original and bandpass filtered signals, as seen on both \texttt{SPEAR} and the baseline \texttt{DAE\_SimNoise}. That being said, \texttt{SPEAR} achieves the lowest error in both HRV metrics out of all the comparison signals. For SDNN, \texttt{SPEAR} achieves an improvement of approximately 60\% over the original signal and 20\% over the denoised signal of \texttt{DAE\_SimNoise}. For RMSSD, \texttt{SPEAR} achieves an improvement of approximately 63\% over the original signal and 32\% over the denoised signal of \texttt{DAE\_SimNoise}. This demonstrates that \texttt{SPEAR}'s denoising algorithm yields significant improvements on the task of HRV estimation from PPG signals.


\begin{table}[h]
\setlength{\tabcolsep}{2pt}
\small
\begin{center}
\begin{tabularx}{\columnwidth}{
  | >{\centering\arraybackslash}X 
  | >{\centering\arraybackslash}X
  | >{\centering\arraybackslash}X
  | >{\centering\arraybackslash}X | }

\hline  
 \textbf{HRV Metric} & Original Signal & Filtered Signal & Denoised (SPEAR) \\
 \hline
 \textbf{SDNN} & 55.98 & 49.93 & 24.16  \\
 \textbf{RMSSD} & 112.47 & 81.19 & 47.7 \\
 \hline
\end{tabularx}
\caption{Mean Absolute Error (MAE) in milliseconds between HRV computed from PPG and ECG Signals. }
\label{HRV-Results}
\end{center}
\end{table}

\section{Discussion}
\subsubsection*{Applications of SPEAR}
PPG technology is becoming increasingly ubiquitous with the adoption of modern wearables such as smartwatches, wristbands and smart jewellery. These devices allow users to continuously monitor heart rate throughout daily life, made possible by their simplicity of operation, cost effectiveness and comfort of use. Most modern wearable devices now also provide continuous HRV measurement \cite{HRVDevices}, which further enables many healthcare applications. PPG has several personal health applications, such as tracking blood pressure \cite{ppgBP} and blood oxygen saturation \cite{ppgapplications}, monitoring sleep quality \cite{ppgsleep}, and guiding exercise and recovery for athletes \cite{fitness-hrv}. Continuous long-term monitoring of PPG also has important clinical applications, such as diagnosing cardiovascular diseases \cite{allen-cardiovascular-ppg} and arrhythmia \cite{Pereira2020Jan}. PPG is highly susceptible to noise and these applications are consequently limited in their accuracy and reliability due to the corruption of underlying metrics obtained from the signals. This is where a denoising algorithm, such as \texttt{SPEAR}, can improve the reliability of these applications. \texttt{SPEAR} can be integrated as a pre-processing step for any application that uses PPG signals for predictive or analytical tasks. The \texttt{SPEAR} algorithm receives a continuous PPG signal of arbitrary duration, splits it into segments, rejects the few segments that are too corrupted to recover and reconstructs the rest only in the noise-corrupted regions, while preserving the useful information in the rest of the signal. The reconstructed clean signal can then be rejoined and passed down to further tasks; since these signals are now clean, they result in more reliable performance in downstream tasks.



\subsubsection*{Deployment} \texttt{SPEAR} can be deployed in personal health applications as well as clinical settings. The algorithm can be implemented in a smartphone application that integrates with the user's wearable. Such an integration only requires a device that provides access to PPG waveform data, such as the Empatica E4 or Actigraph wristwatches (see \cite{HRVDevices} for a review on the data afforded by various PPG enabled devices). For clinical applications, the \texttt{SPEAR} algorithm can be integrated into existing data processing pipelines as a pre-processing step to produce clean PPG signals to be used in various predictive or analytical tasks. \texttt{SPEAR}'s code is \textit{publicly available and open source}; hence, it is free to integrate into any personal or clinical application without requiring FDA approval. This serves three purposes: (1) troubleshooting can be crowdsourced, (2) it can be used as a baseline for comparison with proprietary products, and (3) developers can combine \texttt{SPEAR} with off-the-shelf open source algorithms that work for clean PPG. That said, \texttt{SPEAR} can also be built into algorithms that can apply for FDA approval and have a greater degree of trust and reliability. Most importantly, our work will \textit{provide users the ability to glean metrics on their health, without being limited to the proprietary algorithms provided by device manufacturers}. This can enable the development of a variety of PPG enabled applications available to the public.

\subsubsection*{Code Availability.} Code for \texttt{SPEAR} will be made publicly available at \textcolor{blue}{https://github.com/\{redacted\}}.

\section{Conclusion}
We introduced a novel self-supervised learning paradigm for eliminating noise artifacts and estimating heart rate from PPG signals collected from wrist-worn wearable devices.  An advantage of our approach is that it only requires clean PPG signals for training, which allows us to use larger datasets without ground-truth labels.  \texttt{SPEAR} outperforms baselines at HR and HRV estimation and generalizes well across datasets. This illustrates how \texttt{SPEAR} enables more accurate downstream analysis of many aspects of heart monitoring from wearables.

\bibliographystyle{ACM-Reference-Format}
\bibliography{references}


\begin{thebibliography}{42}


\ifx \showCODEN    \undefined \def \showCODEN     #1{\unskip}     \fi
\ifx \showDOI      \undefined \def \showDOI       #1{#1}\fi
\ifx \showISBNx    \undefined \def \showISBNx     #1{\unskip}     \fi
\ifx \showISBNxiii \undefined \def \showISBNxiii  #1{\unskip}     \fi
\ifx \showISSN     \undefined \def \showISSN      #1{\unskip}     \fi
\ifx \showLCCN     \undefined \def \showLCCN      #1{\unskip}     \fi
\ifx \shownote     \undefined \def \shownote      #1{#1}          \fi
\ifx \showarticletitle \undefined \def \showarticletitle #1{#1}   \fi
\ifx \showURL      \undefined \def \showURL       {\relax}        \fi
\providecommand\bibfield[2]{#2}
\providecommand\bibinfo[2]{#2}
\providecommand\natexlab[1]{#1}
\providecommand\showeprint[2][]{arXiv:#2}

\bibitem[Allen et~al\mbox{.}(2006)]%
        {allen-cardiovascular-ppg}
\bibfield{author}{\bibinfo{person}{John Allen}, \bibinfo{person}{Klaus
  Overbeck}, \bibinfo{person}{Gerard Stansby}, {and} \bibinfo{person}{Alan
  Murray}.} \bibinfo{year}{2006}\natexlab{}.
\newblock \showarticletitle{Photoplethysmography Assessments in Cardiovascular
  Disease}.
\newblock \bibinfo{journal}{\emph{Measurement and Control}}
  \bibinfo{volume}{39}, \bibinfo{number}{3} (\bibinfo{year}{2006}),
  \bibinfo{pages}{80--83}.
\newblock
\urldef\tempurl%
\url{https://doi.org/10.1177/002029400603900303}
\showDOI{\tempurl}
\showeprint{https://doi.org/10.1177/002029400603900303}


\bibitem[Almarshad et~al\mbox{.}(2022)]%
        {ppgapplications}
\bibfield{author}{\bibinfo{person}{Malak~Abdullah Almarshad},
  \bibinfo{person}{Md~Saiful Islam}, \bibinfo{person}{Saad Al-Ahmadi}, {and}
  \bibinfo{person}{Ahmed~S. BaHammam}.} \bibinfo{year}{2022}\natexlab{}.
\newblock \showarticletitle{{Diagnostic Features and Potential Applications of
  PPG Signal in Healthcare: A Systematic Review}}.
\newblock \bibinfo{journal}{\emph{Healthcare}} \bibinfo{volume}{10},
  \bibinfo{number}{3} (\bibinfo{date}{March} \bibinfo{year}{2022}),
  \bibinfo{pages}{547}.
\newblock
\showISSN{2227-9032}
\urldef\tempurl%
\url{https://doi.org/10.3390/healthcare10030547}
\showDOI{\tempurl}


\bibitem[Bartels et~al\mbox{.}(2017)]%
        {SinusCor}
\bibfield{author}{\bibinfo{person}{Rhenan Bartels}, \bibinfo{person}{Leonardo
  Neumamm}, \bibinfo{person}{Tiago Pe{\ifmmode\mbox{\c{c}}\else\c{c}\fi}anha},
  {and} \bibinfo{person}{Alysson Roncally~Silva Carvalho}.}
  \bibinfo{year}{2017}\natexlab{}.
\newblock \showarticletitle{{SinusCor: an advanced tool for heart rate
  variability analysis}}.
\newblock \bibinfo{journal}{\emph{Biomed. Eng. Online}}  \bibinfo{volume}{16}
  (\bibinfo{year}{2017}).
\newblock
\urldef\tempurl%
\url{https://doi.org/10.1186/s12938-017-0401-4}
\showDOI{\tempurl}


\bibitem[Biswas et~al\mbox{.}(2019)]%
        {cornet}
\bibfield{author}{\bibinfo{person}{Dwaipayan Biswas}, \bibinfo{person}{Luke
  Everson}, \bibinfo{person}{Muqing Liu}, \bibinfo{person}{Madhuri Panwar},
  \bibinfo{person}{Bram-Ernst Verhoef}, \bibinfo{person}{Shrishail Patki},
  \bibinfo{person}{Chris~H. Kim}, \bibinfo{person}{Amit Acharyya},
  \bibinfo{person}{Chris Van~Hoof}, \bibinfo{person}{Mario Konijnenburg}, {and}
  \bibinfo{person}{Nick Van~Helleputte}.} \bibinfo{year}{2019}\natexlab{}.
\newblock \showarticletitle{Cor{NET}: Deep Learning Framework for {PPG}-Based
  Heart Rate Estimation and Biometric Identification in Ambulant Environment}.
\newblock \bibinfo{journal}{\emph{IEEE Transac. Biomed. Circuits \& Sys.}}
  \bibinfo{volume}{13}, \bibinfo{number}{2} (\bibinfo{year}{2019}),
  \bibinfo{pages}{282--291}.
\newblock
\urldef\tempurl%
\url{https://doi.org/10.1109/TBCAS.2019.2892297}
\showDOI{\tempurl}


\bibitem[Castaneda et~al\mbox{.}(2018)]%
        {CastanedaPPGApplications}
\bibfield{author}{\bibinfo{person}{Denisse Castaneda}, \bibinfo{person}{Aibhlin
  Esparza}, \bibinfo{person}{Mohammad Ghamari}, \bibinfo{person}{Cinna
  Soltanpur}, {and} \bibinfo{person}{Homer Nazeran}.}
  \bibinfo{year}{2018}\natexlab{}.
\newblock \showarticletitle{{A review on wearable photoplethysmography sensors
  and their potential future applications in health care}}.
\newblock \bibinfo{journal}{\emph{Intl. journal of biosensors {\&}
  bioelectronics}} \bibinfo{volume}{4}, \bibinfo{number}{4}
  (\bibinfo{year}{2018}), \bibinfo{pages}{195}.
\newblock
\urldef\tempurl%
\url{https://doi.org/10.15406/ijbsbe.2018.04.00125}
\showDOI{\tempurl}


\bibitem[Chang et~al\mbox{.}(2021)]%
        {deepheart}
\bibfield{author}{\bibinfo{person}{Xiangmao Chang}, \bibinfo{person}{Gangkai
  Li}, \bibinfo{person}{Guoliang Xing}, \bibinfo{person}{Kun Zhu}, {and}
  \bibinfo{person}{Linlin Tu}.} \bibinfo{year}{2021}\natexlab{}.
\newblock \showarticletitle{{DeepHeart}: A Deep Learning Approach for Accurate
  Heart Rate Estimation from {PPG} Signals}.
\newblock \bibinfo{journal}{\emph{{ACM} Trans. Sen. Netw.}}
  \bibinfo{volume}{17}, \bibinfo{number}{2}, Article \bibinfo{articleno}{14}
  (\bibinfo{date}{jan} \bibinfo{year}{2021}), \bibinfo{numpages}{18}~pages.
\newblock


\bibitem[Charlton(2022)]%
        {rrest-charlton}
\bibfield{author}{\bibinfo{person}{Peter Charlton}.}
  \bibinfo{year}{2022}\natexlab{}.
\newblock \bibinfo{title}{{RRest}}.
\newblock
\newblock
\urldef\tempurl%
\url{https://github.com/peterhcharlton/RRest}
\showURL{%
\tempurl}
\newblock
\shownote{[Online; accessed 13. Sep. 2022]}.


\bibitem[Chong et~al\mbox{.}(2014)]%
        {chong-artifact-detection}
\bibfield{author}{\bibinfo{person}{Jo~Woon Chong}, \bibinfo{person}{Duy~K.
  Dao}, \bibinfo{person}{S.~M.~A. Salehizadeh}, \bibinfo{person}{David~D.
  McManus}, \bibinfo{person}{Chad~E. Darling}, \bibinfo{person}{Ki~H. Chon},
  {and} \bibinfo{person}{Yitzhak Mendelson}.} \bibinfo{year}{2014}\natexlab{}.
\newblock \showarticletitle{{Photoplethysmograph Signal Reconstruction Based on
  a Novel Hybrid Motion Artifact Detection{\textendash}Reduction Approach. Part
  I: Motion and Noise Artifact Detection}}.
\newblock \bibinfo{journal}{\emph{Ann. Biomed. Eng.}} \bibinfo{volume}{42},
  \bibinfo{number}{11} (\bibinfo{date}{Nov.} \bibinfo{year}{2014}),
  \bibinfo{pages}{2238--2250}.
\newblock
\showISSN{1573-9686}
\urldef\tempurl%
\url{https://doi.org/10.1007/s10439-014-1080-y}
\showDOI{\tempurl}


\bibitem[Comtois et~al\mbox{.}(2007)]%
        {RLS-filter}
\bibfield{author}{\bibinfo{person}{Gary Comtois}, \bibinfo{person}{Yitzhak
  Mendelson}, {and} \bibinfo{person}{Piyush Ramuka}.}
  \bibinfo{year}{2007}\natexlab{}.
\newblock \showarticletitle{A Comparative Evaluation of Adaptive Noise
  Cancellation Algorithms for Minimizing Motion Artifacts in a Forehead-Mounted
  Wearable Pulse Oximeter}. In \bibinfo{booktitle}{\emph{IEEE Engg. in Med. and
  Biol. Soc.}} \bibinfo{pages}{1528--1531}.
\newblock
\urldef\tempurl%
\url{https://doi.org/10.1109/IEMBS.2007.4352592}
\showDOI{\tempurl}


\bibitem[Elgendi et~al\mbox{.}(2013)]%
        {elgendi-peak-detection}
\bibfield{author}{\bibinfo{person}{M Elgendi}, \bibinfo{person}{I Norton},
  \bibinfo{person}{M Brearley}, \bibinfo{person}{D Abbott}, {and}
  \bibinfo{person}{D Schuurmans}.} \bibinfo{year}{2013}\natexlab{}.
\newblock \showarticletitle{Systolic Peak Detection in Acceleration
  Photoplethysmograms Measured from Emergency Responders in Tropical
  Conditions}.
\newblock \bibinfo{journal}{\emph{PLoS ONE 8(10): e76585}}
  (\bibinfo{year}{2013}).
\newblock


\bibitem[Galli et~al\mbox{.}(2018)]%
        {kalman}
\bibfield{author}{\bibinfo{person}{Alessandra Galli}, \bibinfo{person}{Claudio
  Narduzzi}, {and} \bibinfo{person}{Giada Giorgi}.}
  \bibinfo{year}{2018}\natexlab{}.
\newblock \showarticletitle{Measuring Heart Rate During Physical Exercise by
  Subspace Decomposition and Kalman Smoothing}.
\newblock \bibinfo{journal}{\emph{IEEE Transactions on Instrumentation and
  Measurement}} \bibinfo{volume}{67}, \bibinfo{number}{5}
  (\bibinfo{year}{2018}), \bibinfo{pages}{1102--1110}.
\newblock
\urldef\tempurl%
\url{https://doi.org/10.1109/TIM.2017.2770818}
\showDOI{\tempurl}


\bibitem[Goh et~al\mbox{.}(2020)]%
        {Goh2020Nov}
\bibfield{author}{\bibinfo{person}{Choon-Hian Goh}, \bibinfo{person}{Li~Kuo
  Tan}, \bibinfo{person}{Nigel~H. Lovell}, \bibinfo{person}{Siew-Cheok Ng},
  \bibinfo{person}{Maw~Pin Tan}, {and} \bibinfo{person}{Einly Lim}.}
  \bibinfo{year}{2020}\natexlab{}.
\newblock \showarticletitle{{Robust {PPG} motion artifact detection using a 1-D
  convolution neural network}}.
\newblock \bibinfo{journal}{\emph{Comput. Methods Programs Biomed.}}
  \bibinfo{volume}{196} (\bibinfo{date}{Nov.} \bibinfo{year}{2020}),
  \bibinfo{pages}{105596}.
\newblock
\showISSN{0169-2607}
\urldef\tempurl%
\url{https://doi.org/10.1016/j.cmpb.2020.105596}
\showDOI{\tempurl}


\bibitem[Guo et~al\mbox{.}(2021)]%
        {segade}
\bibfield{author}{\bibinfo{person}{Zhicheng Guo}, \bibinfo{person}{Cheng Ding},
  \bibinfo{person}{Xiao Hu}, {and} \bibinfo{person}{Cynthia Rudin}.}
  \bibinfo{year}{2021}\natexlab{}.
\newblock \showarticletitle{A supervised machine learning semantic segmentation
  approach for detecting artifacts in plethysmography signals from wearables}.
\newblock \bibinfo{journal}{\emph{Physiological Measurement}}
  \bibinfo{volume}{42}, \bibinfo{number}{12} (\bibinfo{date}{dec}
  \bibinfo{year}{2021}), \bibinfo{pages}{125003}.
\newblock
\urldef\tempurl%
\url{https://doi.org/10.1088/1361-6579/ac3b3d}
\showDOI{\tempurl}


\bibitem[Hinde et~al\mbox{.}(2021)]%
        {HRVDevices}
\bibfield{author}{\bibinfo{person}{Katrina Hinde}, \bibinfo{person}{Graham
  White}, {and} \bibinfo{person}{Nicola Armstrong}.}
  \bibinfo{year}{2021}\natexlab{}.
\newblock \showarticletitle{{Wearable Devices Suitable for Monitoring Twenty
  Four Hour Heart Rate Variability in Military Populations}}.
\newblock \bibinfo{journal}{\emph{Sensors (Basel, Switzerland)}}
  \bibinfo{volume}{21}, \bibinfo{number}{4} (\bibinfo{date}{Feb.}
  \bibinfo{year}{2021}).
\newblock
\urldef\tempurl%
\url{https://doi.org/10.3390/s21041061}
\showDOI{\tempurl}


\bibitem[Kasambe and Rathod(2015)]%
        {vlsi-wavelet}
\bibfield{author}{\bibinfo{person}{P.~V. Kasambe} {and} \bibinfo{person}{S.~S.
  Rathod}.} \bibinfo{year}{2015}\natexlab{}.
\newblock \showarticletitle{{VLSI} Wavelet Based Denoising of {PPG} Signal}.
\newblock \bibinfo{journal}{\emph{Proceedings of ICAC3'15}}
  \bibinfo{volume}{49} (\bibinfo{year}{2015}), \bibinfo{pages}{282--288}.
\newblock
\urldef\tempurl%
\url{https://www.sciencedirect.com/science/article/pii/S1877050915007632}
\showURL{%
\tempurl}


\bibitem[Kingma and Ba(2014)]%
        {adam}
\bibfield{author}{\bibinfo{person}{Diederik~P. Kingma} {and}
  \bibinfo{person}{Jimmy Ba}.} \bibinfo{year}{2014}\natexlab{}.
\newblock \showarticletitle{{Adam: A Method for Stochastic Optimization}}.
\newblock \bibinfo{journal}{\emph{arXiv}} (\bibinfo{date}{Dec.}
  \bibinfo{year}{2014}).
\newblock
\urldef\tempurl%
\url{https://doi.org/10.48550/arXiv.1412.6980}
\showDOI{\tempurl}
\showeprint{1412.6980}


\bibitem[Korkalainen et~al\mbox{.}(2020)]%
        {ppgsleep}
\bibfield{author}{\bibinfo{person}{Henri Korkalainen}, \bibinfo{person}{Juhani
  Aakko}, \bibinfo{person}{Brett Duce}, \bibinfo{person}{Samu Kainulainen},
  \bibinfo{person}{Akseli Leino}, \bibinfo{person}{Sami Nikkonen},
  \bibinfo{person}{Isaac~O Afara}, \bibinfo{person}{Sami Myllymaa},
  \bibinfo{person}{Juha Töyräs}, {and} \bibinfo{person}{Timo Leppänen}.}
  \bibinfo{year}{2020}\natexlab{}.
\newblock \showarticletitle{{Deep learning enables sleep staging from
  photoplethysmogram for patients with suspected sleep apnea}}.
\newblock \bibinfo{journal}{\emph{Sleep}} \bibinfo{volume}{43},
  \bibinfo{number}{11} (\bibinfo{date}{05} \bibinfo{year}{2020}).
\newblock
\showISSN{0161-8105}
\urldef\tempurl%
\url{https://doi.org/10.1093/sleep/zsaa098}
\showDOI{\tempurl}
\showeprint{https://academic.oup.com/sleep/article-pdf/43/11/zsaa098/34499301/zsaa098\_suppl\_supplementary\_material.pdf}
\newblock
\shownote{zsaa098}.


\bibitem[Lam et~al\mbox{.}(2020)]%
        {hrv-real-world}
\bibfield{author}{\bibinfo{person}{Emily Lam}, \bibinfo{person}{Shahrose
  Aratia}, \bibinfo{person}{Julian Wang}, {and} \bibinfo{person}{James Tung}.}
  \bibinfo{year}{2020}\natexlab{}.
\newblock \showarticletitle{Measuring Heart Rate Variability in Free-Living
  Conditions Using Consumer-Grade Photoplethysmography: Validation Study}.
\newblock \bibinfo{journal}{\emph{JMIR Biomed Eng}} \bibinfo{volume}{5},
  \bibinfo{number}{1} (\bibinfo{date}{3 Nov} \bibinfo{year}{2020}),
  \bibinfo{pages}{e17355}.
\newblock
\showISSN{2561-3278}
\urldef\tempurl%
\url{https://doi.org/10.2196/17355}
\showDOI{\tempurl}


\bibitem[Lee et~al\mbox{.}(2019)]%
        {BRDAE}
\bibfield{author}{\bibinfo{person}{Joonnyong Lee}, \bibinfo{person}{Sukkyu
  Sun}, \bibinfo{person}{Seung~Man Yang}, \bibinfo{person}{Jang~Jay Sohn},
  \bibinfo{person}{Jonghyun Park}, \bibinfo{person}{Saram Lee}, {and}
  \bibinfo{person}{Hee~Chan Kim}.} \bibinfo{year}{2019}\natexlab{}.
\newblock \showarticletitle{Bidirectional Recurrent Auto-Encoder for
  Photoplethysmogram Denoising}.
\newblock \bibinfo{journal}{\emph{IEEE Journal of Biomedical and Health
  Informatics}} \bibinfo{volume}{23}, \bibinfo{number}{6}
  (\bibinfo{year}{2019}), \bibinfo{pages}{2375--2385}.
\newblock
\urldef\tempurl%
\url{https://doi.org/10.1109/JBHI.2018.2885139}
\showDOI{\tempurl}


\bibitem[Lu et~al\mbox{.}(2008)]%
        {ppg-hrv}
\bibfield{author}{\bibinfo{person}{Sheng Lu}, \bibinfo{person}{He Zhao},
  \bibinfo{person}{Kihwan Ju}, \bibinfo{person}{Kunson Shin},
  \bibinfo{person}{Myoungho Lee}, \bibinfo{person}{Kirk Shelley}, {and}
  \bibinfo{person}{Ki~H. Chon}.} \bibinfo{year}{2008}\natexlab{}.
\newblock \showarticletitle{{Can Photoplethysmography Variability Serve as an
  Alternative Approach to Obtain Heart Rate Variability Information?}}
\newblock \bibinfo{journal}{\emph{J. Clin. Monit. Comput.}}
  \bibinfo{volume}{22}, \bibinfo{number}{1} (\bibinfo{date}{Jan.}
  \bibinfo{year}{2008}), \bibinfo{pages}{23--29}.
\newblock
\showISSN{1573-2614}
\urldef\tempurl%
\url{https://doi.org/10.1007/s10877-007-9103-y}
\showDOI{\tempurl}


\bibitem[Mishra and Nirala(2020)]%
        {denoise-survey}
\bibfield{author}{\bibinfo{person}{Bhanupriya Mishra} {and}
  \bibinfo{person}{Neelam~Sobha Nirala}.} \bibinfo{year}{2020}\natexlab{}.
\newblock \showarticletitle{A Survey on Denoising Techniques of {PPG} Signal}.
  In \bibinfo{booktitle}{\emph{{IEEE INOCON} Conference}}.
  \bibinfo{pages}{1--8}.
\newblock


\bibitem[Morelli et~al\mbox{.}(2018)]%
        {Morelli2018Apr}
\bibfield{author}{\bibinfo{person}{Davide Morelli}, \bibinfo{person}{Leonardo
  Bartoloni}, \bibinfo{person}{Michele Colombo}, \bibinfo{person}{David Plans},
  {and} \bibinfo{person}{David~A. Clifton}.} \bibinfo{year}{2018}\natexlab{}.
\newblock \showarticletitle{{Profiling the propagation of error from PPG to HRV
  features in a wearable physiological-monitoring device}}.
\newblock \bibinfo{journal}{\emph{Healthcare Technol. Lett.}}
  \bibinfo{volume}{5}, \bibinfo{number}{2} (\bibinfo{date}{April}
  \bibinfo{year}{2018}), \bibinfo{pages}{59--64}.
\newblock
\showISSN{2053-3713}
\urldef\tempurl%
\url{https://doi.org/10.1049/htl.2017.0039}
\showDOI{\tempurl}


\bibitem[Pan et~al\mbox{.}(2016)]%
        {heartBEAT}
\bibfield{author}{\bibinfo{person}{Huijie Pan}, \bibinfo{person}{Dogancan
  Temel}, {and} \bibinfo{person}{Ghassan AlRegib}.}
  \bibinfo{year}{2016}\natexlab{}.
\newblock \showarticletitle{{HeartBEAT}: Heart beat estimation through adaptive
  tracking}. In \bibinfo{booktitle}{\emph{Conference of {BHI}}}.
  \bibinfo{pages}{587--590}.
\newblock


\bibitem[Panwar et~al\mbox{.}(2020)]%
        {PPNet}
\bibfield{author}{\bibinfo{person}{Madhuri Panwar}, \bibinfo{person}{Arvind
  Gautam}, \bibinfo{person}{Dwaipayan Biswas}, {and} \bibinfo{person}{Amit
  Acharyya}.} \bibinfo{year}{2020}\natexlab{}.
\newblock \showarticletitle{{PP}-Net: A Deep Learning Framework for {PPG}-Based
  Blood Pressure and Heart Rate Estimation}.
\newblock \bibinfo{journal}{\emph{IEEE Sensors Journal}} \bibinfo{volume}{20},
  \bibinfo{number}{17} (\bibinfo{year}{2020}), \bibinfo{pages}{10000--10011}.
\newblock
\urldef\tempurl%
\url{https://doi.org/10.1109/JSEN.2020.2990864}
\showDOI{\tempurl}


\bibitem[Peng et~al\mbox{.}(2014)]%
        {cICA}
\bibfield{author}{\bibinfo{person}{Fulai Peng}, \bibinfo{person}{Zhengbo
  Zhang}, \bibinfo{person}{Xiaoming Gou}, \bibinfo{person}{Hongyun Liu}, {and}
  \bibinfo{person}{Weidong Wang}.} \bibinfo{year}{2014}\natexlab{}.
\newblock \showarticletitle{Motion artifact removal from photoplethysmographic
  signals by combining temporally constrained independent component analysis
  and adaptive filter}.
\newblock \bibinfo{journal}{\emph{BioMedical Engineering OnLine}}
  \bibinfo{volume}{13}, \bibinfo{number}{1} (\bibinfo{year}{2014}),
  \bibinfo{pages}{50}.
\newblock


\bibitem[Pereira et~al\mbox{.}(2020)]%
        {Pereira2020Jan}
\bibfield{author}{\bibinfo{person}{Tania Pereira}, \bibinfo{person}{Nate Tran},
  \bibinfo{person}{Kais Gadhoumi}, \bibinfo{person}{Michele~M. Pelter},
  \bibinfo{person}{Duc~H. Do}, \bibinfo{person}{Randall~J. Lee},
  \bibinfo{person}{Rene Colorado}, \bibinfo{person}{Karl Meisel}, {and}
  \bibinfo{person}{Xiao Hu}.} \bibinfo{year}{2020}\natexlab{}.
\newblock \showarticletitle{{Photoplethysmography based atrial fibrillation
  detection: a review}}.
\newblock \bibinfo{journal}{\emph{npj Digital Med.}} \bibinfo{volume}{3},
  \bibinfo{number}{3} (\bibinfo{date}{Jan.} \bibinfo{year}{2020}),
  \bibinfo{pages}{1--12}.
\newblock
\showISSN{2398-6352}
\urldef\tempurl%
\url{https://doi.org/10.1038/s41746-019-0207-9}
\showDOI{\tempurl}


\bibitem[Reiss et~al\mbox{.}(2019)]%
        {deep-ppg}
\bibfield{author}{\bibinfo{person}{Attila Reiss}, \bibinfo{person}{Ina
  Indlekofer}, \bibinfo{person}{Philip Schmidt}, {and} \bibinfo{person}{Kristof
  Van~Laerhoven}.} \bibinfo{year}{2019}\natexlab{}.
\newblock \showarticletitle{Deep {PPG}: Large-Scale Heart Rate Estimation with
  Convolutional Neural Networks}.
\newblock \bibinfo{journal}{\emph{Sensors}} \bibinfo{volume}{19},
  \bibinfo{number}{14} (\bibinfo{year}{2019}).
\newblock


\bibitem[Salehizadeh et~al\mbox{.}(2016)]%
        {SpaMa}
\bibfield{author}{\bibinfo{person}{Seyed M.~A. Salehizadeh},
  \bibinfo{person}{Duy Dao}, \bibinfo{person}{Jeffrey Bolkhovsky},
  \bibinfo{person}{Chae Cho}, \bibinfo{person}{Yitzhak Mendelson}, {and}
  \bibinfo{person}{Ki~H. Chon}.} \bibinfo{year}{2016}\natexlab{}.
\newblock \showarticletitle{A Novel Time-Varying Spectral Filtering Algorithm
  for Reconstruction of Motion Artifact Corrupted Heart Rate Signals During
  Intense Physical Activities Using a Wearable Photoplethysmogram Sensor}.
\newblock \bibinfo{journal}{\emph{Sensors}} \bibinfo{volume}{16},
  \bibinfo{number}{1} (\bibinfo{year}{2016}).
\newblock
\showISSN{1424-8220}
\urldef\tempurl%
\url{https://doi.org/10.3390/s16010010}
\showDOI{\tempurl}


\bibitem[Sa{\ifmmode\tilde{n}\else\~{n}\fi}udo et~al\mbox{.}(2019)]%
        {sanudo}
\bibfield{author}{\bibinfo{person}{Borja
  Sa{\ifmmode\tilde{n}\else\~{n}\fi}udo},
  \bibinfo{person}{Mois{\ifmmode\acute{e}\else\'{e}\fi}s De~Hoyo},
  \bibinfo{person}{Alejandro
  Mu{\ifmmode\tilde{n}\else\~{n}\fi}oz-L{\ifmmode\acute{o}\else\'{o}\fi}pez},
  \bibinfo{person}{John Perry}, {and} \bibinfo{person}{Grant Abt}.}
  \bibinfo{year}{2019}\natexlab{}.
\newblock \showarticletitle{{Pilot Study Assessing the Influence of Skin Type
  on the Heart Rate Measurements Obtained by Photoplethysmography with the
  Apple Watch}}.
\newblock \bibinfo{journal}{\emph{J. Med. Syst.}} \bibinfo{volume}{43},
  \bibinfo{number}{7} (\bibinfo{date}{July} \bibinfo{year}{2019}),
  \bibinfo{pages}{1--8}.
\newblock
\showISSN{1573-689X}
\urldef\tempurl%
\url{https://doi.org/10.1007/s10916-019-1325-2}
\showDOI{\tempurl}


\bibitem[Schäck et~al\mbox{.}(2015)]%
        {shaeck-hr}
\bibfield{author}{\bibinfo{person}{Tim Schäck}, \bibinfo{person}{Christian
  Sledz}, \bibinfo{person}{Michael Muma}, {and} \bibinfo{person}{Abdelhak~M.
  Zoubir}.} \bibinfo{year}{2015}\natexlab{}.
\newblock \showarticletitle{A new method for heart rate monitoring during
  physical exercise using photoplethysmographic signals}. In
  \bibinfo{booktitle}{\emph{{EUSIPCO} Conference}}.
\newblock
\urldef\tempurl%
\url{https://doi.org/10.1109/EUSIPCO.2015.7362868}
\showDOI{\tempurl}


\bibitem[Shaffer and Ginsberg(2017)]%
        {hrv-overview-shaffer}
\bibfield{author}{\bibinfo{person}{Fred Shaffer} {and} \bibinfo{person}{J~P
  Ginsberg}.} \bibinfo{year}{2017}\natexlab{}.
\newblock \showarticletitle{An Overview of Heart Rate Variability Metrics and
  Norms}.
\newblock \bibinfo{journal}{\emph{Frontiers in public health}}
  \bibinfo{volume}{5} (\bibinfo{date}{09} \bibinfo{year}{2017}),
  \bibinfo{pages}{258--258}.
\newblock
\showISBNx{2296-2565; 2296-2565}
\urldef\tempurl%
\url{https://doi.org/10.3389/fpubh.2017.00258}
\showDOI{\tempurl}


\bibitem[Shyam et~al\mbox{.}(2019)]%
        {ppgnet}
\bibfield{author}{\bibinfo{person}{A. Shyam}, \bibinfo{person}{Vignesh
  Ravichandran}, \bibinfo{person}{S.P. Preejith}, \bibinfo{person}{Jayaraj
  Joseph}, {and} \bibinfo{person}{Mohanasankar Sivaprakasam}.}
  \bibinfo{year}{2019}\natexlab{}.
\newblock \showarticletitle{{PPG}net: Deep Network for Device Independent Heart
  Rate Estimation from Photoplethysmogram}. In \bibinfo{booktitle}{\emph{Intl.
  Conference of {IEEE EMBC}}}. \bibinfo{pages}{1899--1902}.
\newblock
\urldef\tempurl%
\url{https://doi.org/10.1109/EMBC.2019.8856989}
\showDOI{\tempurl}


\bibitem[Singh et~al\mbox{.}(2018)]%
        {fitness-hrv}
\bibfield{author}{\bibinfo{person}{Nikhil Singh}, \bibinfo{person}{Kegan~James
  Moneghetti}, \bibinfo{person}{Jeffrey~Wilcox Christle},
  \bibinfo{person}{David Hadley}, \bibinfo{person}{Victor Froelicher}, {and}
  \bibinfo{person}{Daniel Plews}.} \bibinfo{year}{2018}\natexlab{}.
\newblock \showarticletitle{{Heart Rate Variability: An Old Metric with New
  Meaning in the Era of Using mHealth technologies for Health and Exercise
  Training Guidance. Part Two: Prognosis and Training}}.
\newblock \bibinfo{journal}{\emph{Arrhythmia Electrophysiol. Rev.}}
  \bibinfo{volume}{7}, \bibinfo{number}{4} (\bibinfo{date}{Dec.}
  \bibinfo{year}{2018}), \bibinfo{pages}{247}.
\newblock
\urldef\tempurl%
\url{https://doi.org/10.15420/aer.2018.30.2}
\showDOI{\tempurl}


\bibitem[Temko(2017)]%
        {temko}
\bibfield{author}{\bibinfo{person}{Andriy Temko}.}
  \bibinfo{year}{2017}\natexlab{}.
\newblock \showarticletitle{{PPG}-based heart rate estimation using Wiener
  filter, phase vocoder and Viterbi decoding}. In
  \bibinfo{booktitle}{\emph{{IEEE} {ICASSP}}}. \bibinfo{pages}{1013--1017}.
\newblock
\urldef\tempurl%
\url{https://doi.org/10.1109/ICASSP.2017.7952309}
\showDOI{\tempurl}


\bibitem[Torres-Soto and Ashley(2020)]%
        {stanford-data}
\bibfield{author}{\bibinfo{person}{Jessica Torres-Soto} {and}
  \bibinfo{person}{Euan~A. Ashley}.} \bibinfo{year}{2020}\natexlab{}.
\newblock \showarticletitle{{Multi-task deep learning for cardiac rhythm
  detection in wearable devices}}.
\newblock \bibinfo{journal}{\emph{npj Digital Med.}} \bibinfo{volume}{3},
  \bibinfo{number}{116} (\bibinfo{date}{Sept.} \bibinfo{year}{2020}),
  \bibinfo{pages}{1--8}.
\newblock
\showISSN{2398-6352}
\urldef\tempurl%
\url{https://doi.org/10.1038/s41746-020-00320-4}
\showDOI{\tempurl}


\bibitem[Tsipouras and Fotiadis(2004)]%
        {arrhythmia-hrv}
\bibfield{author}{\bibinfo{person}{Markos~G. Tsipouras} {and}
  \bibinfo{person}{Dimitrios~I. Fotiadis}.} \bibinfo{year}{2004}\natexlab{}.
\newblock \showarticletitle{{Automatic arrhythmia detection based on time and
  time{\textendash}frequency analysis of heart rate variability}}.
\newblock \bibinfo{journal}{\emph{Comput. Methods Programs Biomed.}}
  \bibinfo{volume}{74}, \bibinfo{number}{2} (\bibinfo{date}{May}
  \bibinfo{year}{2004}), \bibinfo{pages}{95--108}.
\newblock
\showISSN{0169-2607}
\urldef\tempurl%
\url{https://doi.org/10.1016/S0169-2607(03)00079-8}
\showDOI{\tempurl}


\bibitem[Tsuji et~al\mbox{.}(1996)]%
        {stroke-hrv}
\bibfield{author}{\bibinfo{person}{Hisako Tsuji}, \bibinfo{person}{Martin~G.
  Larson}, \bibinfo{person}{Ferdinand~J. Venditti}, \bibinfo{person}{Emily~S.
  Manders}, \bibinfo{person}{Jane~C. Evans}, \bibinfo{person}{Charles~L.
  Feldman}, {and} \bibinfo{person}{Daniel Levy}.}
  \bibinfo{year}{1996}\natexlab{}.
\newblock \showarticletitle{{Impact of Reduced Heart Rate Variability on Risk
  for Cardiac Events}}.
\newblock \bibinfo{journal}{\emph{Circulation}} (\bibinfo{date}{Dec.}
  \bibinfo{year}{1996}).
\newblock
\urldef\tempurl%
\url{https://doi.org/10.1161/01.CIR.94.11.2850}
\showDOI{\tempurl}


\bibitem[van Gent et~al\mbox{.}(2018)]%
        {heartpy}
\bibfield{author}{\bibinfo{person}{Paul van Gent}, \bibinfo{person}{Haneen
  Farah}, \bibinfo{person}{Nicole Nes}, {and} \bibinfo{person}{B. Arem}.}
  \bibinfo{year}{2018}\natexlab{}.
\newblock \showarticletitle{Heart Rate Analysis for Human Factors: Development
  and Validation of an Open Source Toolkit for Noisy Naturalistic Heart Rate
  Data}.
\newblock


\bibitem[Vincent et~al\mbox{.}(2008)]%
        {denoise-ae}
\bibfield{author}{\bibinfo{person}{Pascal Vincent}, \bibinfo{person}{Hugo
  Larochelle}, \bibinfo{person}{Yoshua Bengio}, {and}
  \bibinfo{person}{Pierre-Antoine Manzagol}.} \bibinfo{year}{2008}\natexlab{}.
\newblock \showarticletitle{Extracting and Composing Robust Features with
  Denoising Autoencoders}. In \bibinfo{booktitle}{\emph{Proceedings of
  {ICML}}}.
\newblock
\showISBNx{9781605582054}
\urldef\tempurl%
\url{https://doi.org/10.1145/1390156.1390294}
\showDOI{\tempurl}


\bibitem[Wu et~al\mbox{.}(2017)]%
        {adaptive-filter-wu}
\bibfield{author}{\bibinfo{person}{Chih-Chin Wu}, \bibinfo{person}{I-Wei Chen},
  {and} \bibinfo{person}{Wai-Chi Fang}.} \bibinfo{year}{2017}\natexlab{}.
\newblock \showarticletitle{An implementation of motion artifacts elimination
  for {PPG} signal processing based on recursive least squares adaptive
  filter}. In \bibinfo{booktitle}{\emph{{IEEE BioCAS} Conference}}.
  \bibinfo{pages}{1--4}.
\newblock
\urldef\tempurl%
\url{https://doi.org/10.1109/BIOCAS.2017.8325141}
\showDOI{\tempurl}


\bibitem[Yoon et~al\mbox{.}(2009)]%
        {ppgBP}
\bibfield{author}{\bibinfo{person}{Youngzoon Yoon}, \bibinfo{person}{Jung~H.
  Cho}, {and} \bibinfo{person}{Gilwon Yoon}.} \bibinfo{year}{2009}\natexlab{}.
\newblock \showarticletitle{{Non-constrained Blood Pressure Monitoring Using
  ECG and PPG for Personal Healthcare}}.
\newblock \bibinfo{journal}{\emph{J. Med. Syst.}} \bibinfo{volume}{33},
  \bibinfo{number}{4} (\bibinfo{date}{Aug.} \bibinfo{year}{2009}),
  \bibinfo{pages}{261--266}.
\newblock
\showISSN{1573-689X}
\urldef\tempurl%
\url{https://doi.org/10.1007/s10916-008-9186-0}
\showDOI{\tempurl}


\bibitem[Zhang et~al\mbox{.}(2015)]%
        {troika}
\bibfield{author}{\bibinfo{person}{Zhilin Zhang}, \bibinfo{person}{Zhouyue Pi},
  {and} \bibinfo{person}{Benyuan Liu}.} \bibinfo{year}{2015}\natexlab{}.
\newblock \showarticletitle{{TROIKA}: A General Framework for Heart Rate
  Monitoring Using Wrist-Type Photoplethysmographic Signals During Intensive
  Physical Exercise}.
\newblock \bibinfo{journal}{\emph{IEEE Transactions on Biomedical Engineering}}
  \bibinfo{volume}{62}, \bibinfo{number}{2} (\bibinfo{year}{2015}),
  \bibinfo{pages}{522--531}.
\newblock
\urldef\tempurl%
\url{https://doi.org/10.1109/TBME.2014.2359372}
\showDOI{\tempurl}


\end{thebibliography}

\end{document}